# Natural Selection as the Sum over All Histories

**Abstract:** If Evolution can be connected to the principle of least action, and if it is depicted in evolution space versus time" then it corresponds to the direction of ultimate causation. As an organism evolves and follows a path of proximate causation, if the vector is closely parallel to that of the Ultimate Causation then the changes will confer desirable attributes which will lead to further development. If however, the variations do not occur in a direction close to the of the ultimate causation vector the evolved organism will quickly die out. This may be viewed as similar to Feynman's sum over all histories. Therefore, the principle of least action gives a direction, but not a purpose, to evolution. Taking the coevolution model of Lewontin, an equation of motion for coevolution shows that it is the rate of evolutionary change that responds to changes in the environment, in line with some evidence. In the face of widely held views on mass extinctions because of climate change this gives some small hope.

**Background:**

Since the publication of "On the Origin of Species by Natural Selection" by Charles Darwin in 1859, there have been a number of attempts to link it to other scientific principles, notably the Principle of stationary Action, known popularly as least Action [Nahin]. In their paper Natural selection for least action (Kaila and Annila 2008) Kaila and Annila depict evolution as a process conforming to the Principle of Least Action (PLA). This paper, although not giving any experimental evidence, shows that evolution, if conforming to the Second Law of Thermodynamics, will follow a trajectory of maximum Entropy production, which conforms to the PLA. To demonstrate this they rewrote the Gibbs-Duhem relationship in terms of possible states, to give a differential equation of evolution. This is a convincing argument as biology at root is a physical process, as expounded by Schrodinger [What is Life], and all physical processes are governed by the Second Law of thermodynamics.

The PLA first arose, in modern times in the 17$^{th}$ century, and was popularised by Maupertius [Kaila & Annila] although the major rigorous work was done by Euler [Kaila & Annila]. It is the general principle that transcends all others, and is a Grand Universal Theory of Everything (GUT) at the level of Energy; it was Feynman's guiding principle throughout this life. As Wittgenstein [Tractatus Logicus Philiosphicus] said "people were using the Principle of Least Action before they knew it existed".

The Principle of Least (Stationary) Action is stated simply as:

$$S = \int_0^t U.dt \ \text{and} \ \frac{d}{dt}[S]=0 \qquad (1)$$

Where U is the Gibb's Free Energy.

The second law can be stated as:

$$F = E - E_{nt} \qquad (2)$$

Where: E is the total energy of the body, F is the "Gibbs Free Energy" and $E_{nt}$ is the entropy.

It simply states that in all processes energy is conserved, and the change of entropy (work) has to be maximised if the entropy is to be minimised.

This needs careful interpretation in terms of a body (whatever it is). If the body wishes to extract the least amount of energy from its surroundings (external entropy), the internal processes must be at their most efficient (maximum internal entropy production). This is most clearly seen in a single crystal of material undergoing tensile deformation, when the slip planes orientate themselves to be parallel to the applied force, thereby maximising the amount of internal deformation for a given external force. This is known as Cottrell Rotation [Sills & Cai]. It is also why for a given amount of deformation the force vector is ideally parallel to the deformation vector, to give the most deformation (work done) for a given force; the same principle also applies to moments.

**Results:**

Here it is hypothesised that an analogue exists to Feynman's "Sum Over All Histories" (Feynman 1985). Richard Feynman, Nobel Laureate, one of the greatest minds of the 20[th] century, was fascinated by the PLA for most of his life, and it led to possibly one of his greatest breakthroughs in Quantum Electrodynamics or QED – the quantum theory of light [QED Feynman 1985]. Feynman considered that a Photon traversing a path from A to B in space-time was free to traverse any path. However, when the sum of all paths is taken, because of phase differences, the most likely path is that conforming to the PLA; this is known as the "Sum Over All Histories".

It is assumed here that there is a general overall direction of evolution. The vector of evolution is called here the Ultimate Causal Direction (green vector), in accordance with the terminology used in evolutionary biology(Mayr 2001), and this should conform to the PLA [Kaila & Annil], in conformity with the external conditions, and could be shown graphically as shown in Fig.1.

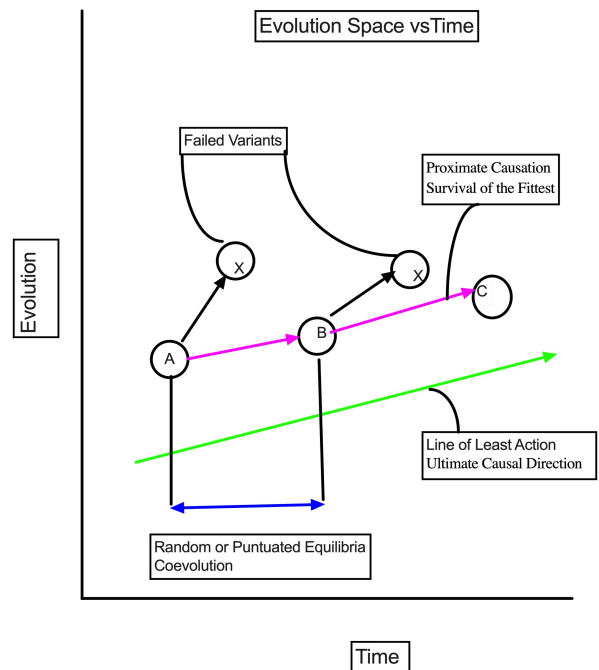

Fig.1 Evolution-Time Space.

If during any genetic mutation leading to a change in the organism takes place (event A) this will in general be random, under constraints [Holmes] and confer on the next generation of organism's thermodynamic attributes which may or may not lead to a greater probability of survival in the external climate.  These changes to the phenotype are in evolutionary biological terms called "Proximate Causation".  It is proposed that Proximate Causation will lead to greater survivability if it is parallel, or closely parallel, to the direction of PLA, for the external conditions at the time (fitness). This is shown in Fig.1. as the red vector A-B-C. In this diagram, those variants not reasonably parallel to the trajectory of ultimate causality, those black lines ending in blobs, will die out. Whereas those variants which are closely parallel to the trajectory of the environment, denoted by red arrows, will be heritable and survive. Further Epigenetic changes [Dupont] could also more closely align the red vector with the Ultimate Causality. Although this diagram as an axis of Time, the interval between the changes is not known.

The eventual unfolding of Proximate Causation towards Ultimate Causation as determined by both thermodynamics and the environment can be considered as analogous to Feynman's "Sum over all Histories", as the mutations leading to Proximate causation may be viewed as exploring all the possible states in evolution space, but only those approximately parallel to the environment vector will survive to generate new variants and lead to the Ultimate Causality.  The other variants will die out. Hence evolution towards the Ultimate Causality can be seen as the ultimate goal of the sum over all proximate evolutionary histories.

This is not to say the evolution to the PLA is teleological, there is no "purpose". It is simply that those evolutionary lines arising from proximate causality that are closely parallel to the line of PLA, are best adapted to whatever external environments that exist at the time, and so survive longer, and have a greater probability of reproduction (heritable); therefore, evolution has a direction but not a purpose.

There are also interesting developments in evolutionary thinking where some scientists are beginning to posit that organisms make changes to their phenotype, as opposed to their genotype, by adjusting to their environment (Holmes, New Scientist).  This tendency here would be to add a rotation to the Proximate Causality, hence bringing the vector closer to the direction of the Ultimate Causality, in the direction of PLA. Hence again, Evolution does not have a purpose but it has a direction.

It was difficult to reconcile the original Gaia hypothesis [Lovelock 1979], however later variants called "Coevolution" [Lewontin] allows reconciliation with Gaia. Lewontin's description of Coevolution as two, coupled, differential equations to express the interaction between the organism (O) and the environment (E) as:

$$\frac{d}{dt}(O) = f(O, E) \tag{3}$$

and $$\frac{d}{dt}(E) = g(O, E) \tag{4}$$

is an interesting development. Taking Kaila and Annila 2008 idea that the second law can be rewritten as an equation of motion, and combining this with Lagrangian's gives;

$$\mathcal{L}_O = f\{f(O, E)\} \tag{5}$$

$$\mathcal{L}_E = f\{g(O, E)\} \tag{6}$$

Using Euler-Lagrange gives:

$$\frac{d}{dt}\mathcal{L}_O f\{f(O,E)\} - \mathcal{L}_E f\{g(O,E)\} = 0 \qquad (7)$$

Hence:

$$\frac{d}{dt}\mathcal{L}_O f\{f(O,E)\} = \mathcal{L}_E f\{g(O,E)\} \qquad (8)$$

Therefore this equation gives an equation of motion for Coevolution and shows that the rate of organism change (rate of evolution) is proportional to the magnitude of the change occurring in the environment, which is in agreement with some evidence [Husby]. This is important as it shows that it is not evolution which is driven by changes in the environment , evolution by mutation can occur at any time, but it is the rate of evolution which changes. In the face of widely held views on mass extinctions because of climate change this gives some small hope [Husby].

There is a measurement problem in evolution through the fossil record in that the rate of progress is inversely proportional to the timescale over which it is measured, leading to Gould's "Punctuated Equilibrium" ideas (Gould and Eldredge 1993). Furthermore, another measurement problem is that of actual findings and observation. In general the probability of finding a fossil will be proportional to the number of the species that existed, and so the time the fossil type lived. So the probability of finding a fossil type is dependent on how successful that fossil type was, and so most of the changes that have occurred in a species will probably never be found because they died out quickly. This is an exceptionally complicated problem, said by [Loewe and Hill] that it is "like looking for a needle in a haystack', however, advances in Evolutionary Genetics may give answers in the future [Loewe and Hill].

**Conclusions:**

It is shown that if Evolution can be connected to the Principle of Least Action, which has been theorised on for many years, and if it is depicted in "Evolution Space vs Time" then it corresponds to the direction of Ultimate Causation. As an organism evolves and follows a path of Proximate Causation, if the vector is closely parallel to that of the Ultimate Causation then the changes will confer desirable heritable attributes which will lead to further development. If however the variations do not occur in a direction close to the of the Ultimate Causation vector the evolved organism will quickly die out. This may be viewed as similar to Feynman's "sum over all histories. Therefore, the Principle of Least Action gives a direction, but not a purpose, to Evolution. Taking the Coevolution model of Lewontin, an equation of motion derivedfor Coevolution shows that it is the rate of evolutionary change that responds to changes in the environment. In the face of widely held views on mass extinctions because of climate change this gives some small hope.

**References:**

Darwin C, On the Origin of Species by Natural Selection, John Murray, London, 1859.

Nahin, PJ, When Least is Best, Princeton UP. 2004.

Kaila, V. R. I. and A. Annila (2008). "Natural selection for least action." Proc.R. Soc A(464): 3055-3070.

Tractatus Logicus-Philosophicus, L Wittgenstein, Routledge 1922.

Schrodinger, E. (1944). What is life?, Cambridge University Press 2007

Feynman, R. P. (1985). QED: The Starnge Theory of Light and Matter, Penguin.
Mayr, E. (2001). What Evolution Is, Basic Books.

Free energy change of a dislocation due to a Cottrell atmosphere. R. B. Sills & W. Cai Phil. Mag Pages 1491-1510 | 07 Mar 2018

Holmes , Life's Purpose, New Scientist, p33, 12 October 2013.

Dupont C, Armant DR, Brenner CA (September 2009). "Epigenetics: definition, mechanisms and clinical perspective". Seminars in Reproductive Medicine. **27** (5): 351–7.

Gaia, J Lovelock, 1979, OUP

Lewontin, R.C. (1983) Gene, organism and environment. In Evolution
from Molecules to Men (Bendall, D.S., ed.), pp. 273–285, Cambridge

Gould, S. J. and N. Eldredge (1993). "Punctuated Equilibrium Comes of Age." Nature **366**: 223-227.
          the
Husby A, Visser ME, Kruuk LEB (2011) Speeding up Microevolution: The Effects of increasing Temperature on Selection and Genetic Variance in a Wild Bird Population. doi:10.1371/journal.pbio.1000585

Free A,2 and Barton NH, Do evolution and ecology need the Gaia hypothesis?
TRENDS in Ecology and Evolution Vol.22 No.11

Loewe L and Hill WG, The population genetics of mutations: good, bad and indifferent; Phil. Trans. R. Soc. B (2010) 365, 1153–1167
doi:10.1098/rstb.2009.0317